\documentclass[useAMS,usenatbib,usegraphicx]{mn2e}

\def\kms{km s$^{-1}$}
\def\hi{H{\sc i}}
\def\hii{H{\sc ii}}
\def\msun{M$_\odot$}

\def\msunyr{M$_\odot$ yr$^{-1}$}
\def\cmdos{cm$^{-2}$}

\def\por{$\times$}
\def\deg{$^\circ$}

\title[\hi\ bubble around WR\,85 and RCW\,118]{An \hi\ interstellar bubble 
surrounding WR\,85 and RCW\,118}

\author[J. Vasquez, C. Cappa and N. McClure-Griffiths]{J. 
Vasquez$^{1}$\thanks{Fellow of CONICET, Argentina\
E-mail: pete@tux.iar.unlp.edu.ar}, 
C. Cappa$^{1,2}$\thanks{Member of Carrera del Investigador, CONICET, 
Argentina}
and N.M. McClure-Griffiths$^{3}$\\
$^{1}$Instituto Argentino de Radioastronom\'{\i}a, C.C.5. 1894, Villa Elisa, 
Argentina\\
$^{2}$Facultad de Ciencias Astron\'omicas y Geof\'{\i}sicas, Universidad 
Nacional de La Plata, La Plata, Argentina\\
$^{3}$Australia Telescope National Facility, CSIRO; P.O. Box 76, Epping NSW 
 1710, Australia}

\begin{document}

\date{Accepted 2005 June 23, Received 2005 June 22; in original form 2005 March 23}

\pagerange{\pageref{firstpage}--\pageref{lastpage}} \pubyear{2002}

\maketitle

\label{firstpage}

\begin{abstract}
We analyze the distribution of the interstellar matter in the environs  of 
the Wolf-Rayet star LSS\,3982 (= WR\,85, WN6+OB?) linked to the optical 
ring nebula RCW\,118. Our study is based on neutral hydrogen 21cm-line data 
belonging to the Southern Galactic Plane Survey (SGPS).\

The analysis of the \hi\ data allowed the identification of a neutral
hydrogen interstellar bubble related to WR\,85 and the 25\arcmin -diameter
ring nebula RCW\,118. 
The \hi\ bubble was detected at a systemic velocity of --21.5 \kms,
corresponding to a kinematical distance of 2.8$\pm$1.1 kpc, compatible
with the stellar distance. The neutral stucture is about 25$^{'}$ in 
radius or 21$\pm$8 pc, and is expanding at 9$\pm$2 \kms. 
The associated ionized and neutral masses amount to 3000 $M_{\odot}$. 
The CO emission distribution depicts a region lacking CO coincident in position
and velocity with the \hi\ structure.
 The 9\farcm 3-diameter inner optical nebula appears to be related to the
approaching part of the neutral atomic shell.
The \hi\ void and shell are the neutral gas counterparts of the optical bubble
and have very probably originated in the action of the strong stellar wind of the central
star during the O-type and WR phases on the surrounding 
interstellar medium. 
The \hi\ bubble appears to be in the momentun conserving stage. 
\end{abstract}

\begin{keywords}
ISM:\ bubbles -- stars: Wolf-Rayet -- ISM:\ \hii\ regions
\end{keywords}

\section{Introduction}

Wolf-Rayet (WR) stars are the evolutionary descendents of massive O-type
stars. With mass loss rates in the interval (1-5)\por 10$^{-5}$ \msunyr\ and 
terminal velocities of 1000\,-\,3000 \kms\ (\citealt{vh}; \citealt{ca3}), 
these hot and luminous stars are one of the most powerful stellar wind 
sources in our Galaxy. Of-type stars are also characterized by high
mass loss rates and terminal velocities (Lamers \& Leitherer 1993; Prinja, 
Barlow \& Howarth 1990).

The mechanical energy released to the interstellar medium (ISM) during the 
WR phase only ($t_{WR} \lse$ 7\por 10$^5$ yr, \citealt{mm})
is in the range (1-30)\por 10$^{50}$ erg, comparable to the mechanical energy 
injected during a supernova explosion. Both the mass flow from the
WR star and the previous O-type star phases strongly modify 
the energetics, the morphology and the chemical abundances of the ISM in the
environs of the star.

The interaction of the strong stellar winds with the surrounding interestellar
matter has been analyzed by several authors, taking into account different 
enviroments (e.g. \citealt{gs}). The stellar flow 
sweeps up the surrounding material creating {\it interstellar bubbles}, 
which are detected at different wavelengths from the UV to the radio range.
In the optical regime, these structures are generally observed as 
filamentary ring nebulae in the light of H$\alpha$ and [O{\sc iii}] (e.g. 
\citealt{chu1}; \citealt{ma2}). They are related to many WR stars and 
to a relatively large number of O and Of stars. Shell-shaped structures 
created by stellar winds from massive stars are also 
identified in the far infrared and in the thermal radio 
continuum emission (e.g. \citealt{mat}; Cappa, Goss \& Pineault 2002).

Interstellar bubbles appear as cavities and expanding shells in the 
neutral hydrogen 21cm line emission distribution (e.g. \citealt{ca1}
and references therein). \hi\ bubbles are external
to their optical and radio continuum counterparts,  and expand at relatively
low velocities ($\lse$ 10 \kms). In most of the cases the 
derived dynamical ages are
larger than the duration of the WR phase, suggesting that 
the stellar winds of the massive O-type star progenitor has also 
contributed to creating the structures.
 
As part of a systematic search for neutral gas bubbles around massive stars, 
we present here a study of the ISM surrounding WR\,85 (= HD\,155603B = 
LSS\,3982) based mainly on  
\hi\ 21cm-line  data belonging to the Southern Galactic Plane Survey 
(SGPS) and additional molecular and radio continuum data.

WR\,85 is classified as WN6h\,+\,OB?. 
Different distance estimates have been published for this star. WR\,85 
belongs to the HD\,155603 group, for which \citet{mf} 
find a distance of 1.8 kpc. \citet{cv} and \citet{vh} 
locate the star at d = 3.7$\pm$1.3 and 4.7$\pm$2.3 kpc, respectively,
while Hipparcos measurements indicate 1.7 kpc. The X-ray point like 
source 1\,WGA\,J1714.4-3949 coincides in position with the WR star 
\citep{pa}.

HD\,155603B is associated with the optical ring nebula RCW\,118, of about 
25$^{'}$ in diameter (\citealt{ct}; Heckathorn, Bruhweiler \& Gull 1982, 
\citealt{ma1}; \citealt{ma2}b). \citealt{chu1}
observed RCW\,118 with the Curtis-Schmidt-0.6m telescope at CTIO in H$\alpha$, 
[O{\sc iii}]$\lambda$5007 and [S{\sc ii}]$\lambda$6730.
These authors found that the ionized gas
in the nebula has a LSR velocity of --15 \kms, and  derived an expansion velocity $\lid$ 10 \kms\ and a 
dynamical age of 6.7\por 10$^5$ yr, compatible with the duration of the 
WR phase. Based on the velocity of the ionized gas and on standard galactic 
rotation models, they estimated a kinematical distance d$_k \sim$ 2.3 kpc.

\begin{figure}
   \includegraphics[angle=0,width=8.4cm,height=7.5cm]{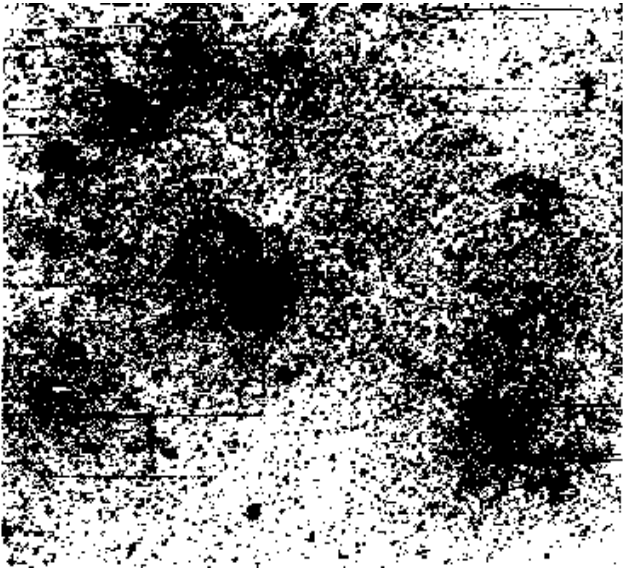}
  \caption{H$\alpha$ image in ($\alpha$, $\delta$) coordinates showing 
the two concentic rings around WR\,85 (taken from \citealt{ma2}). North is up and 
East is to the left.}
  \label{optica}
\end{figure}

\citet{ma2} reobserved the nebula using the same telescope. 
Their H$\alpha$ image reveals the presence of two semi-circular
rings. The inner ring has a diameter of 9\farcm 3 and is centered on the
WR star, whereas the outer one, 25\arcmin\ in diameter,  is centered 
slightly to the south and south-east of the star. The two concentric 
structures can be clearly identified in the H$\alpha$ image displayed in
Figure~\ref{optica}. 

RCW\,118 is in the same line of sight to the SNR G347.3--0.5, discovered 
in the {\sc rosat} All Sky Survey (RASS) by \citet{pa}.
This SNR, 1$^{\circ}$ in diameter, is located at $\approx$ 6 kpc 
(see \citealt{la}).

\section[]{Databases}

The \hi\ 21cm-line emission data used in this paper belong to the 
Southern Galactic Plane Survey (SGPS) (\citealt{McClure})
 obtained with the Autralia Telescope
Compact Array (ATCA) and the Parkes radiotelescope (short spacing 
information). The \hi\ data cube is centered at ($l,b,v$) = 
(347$^{\circ}$14\arcmin,--0$^{\circ}$29\arcmin,--40 \kms), covers a 
region of about 2$^{\circ}$\por 2$^{\circ}$ around the WR star, and has a 
synthesized beam of 2\farcm 6\por 2\farcm 1. To improve the S/N ratio we 
applied a Hanning smoothing to the individual line images. Consequently,
the original rms noise level of 2.4 K was lowered to 1.3 K and the channel 
velocity resolution was doubled. The main observational parameters of the 
final data cube are summarized in Table~\ref{tabla1}.

\begin{table}
 \begin{minipage}{84mm}
\caption{\hi\ data: main observational parameters}\label{tabla1}
\begin{tabular}{l c}
\hline 
($l$,$b$) center       & 347$^{\circ}$14\arcmin, --0$^{\circ}$29\arcmin  \\
Velocity range         &  --204, 124 \kms \\
Velocity resolution     & 1.64 \kms  \\
Surveyed area          & 2$^{\circ}$\por 2$^{\circ}$ \\
{\sc rms}  noise       &  1.3 K     \\
Synthesized beam       & 2\farcm 6\por 2\farcm 1 \\
\hline
\end{tabular}
\end{minipage}
\end{table}

Additional infrared, radio continuum and molecular data were also analyzed. 
High resolution infrared images (HIRES) were obtained from IPAC\footnote{IPAC 
is fouded
by NASA as part of the IRAS extended mission under contract to Jet Propulsion
Laboratory (JPL) and California Intitute of Technology (Caltech).}. 
The {\sc iras} data, obtained at 12, 25, 60 and 100$\mu$m, have angular 
resolutions in the range 0\farcm 5 to about 2\arcmin.
Radio continuum data at 2.4 GHz were obtained from the survey by \citet{du} 
with an angular resolution of 10\farcm 4.
The molecular data corresponds to the CO (J = 1$\rightarrow$0) line
at 115 GHz and  belong to the CO survey by \citet{da}, with angular 
and velocity resolutions of 9\arcmin and 1.3 \kms, respectively, and
a rms noise of 0.3 K.

\section[]{Ionized and neutral gas distribution towards WR\,85}

\subsection[]{\hi\ line emission distribution}

The strong stellar winds from massive stars are expected to sweep-up the
interstellar material around the wind source and to create a highly evacuated
region surrounded by an expanding shell. If the ionizing front is 
trapped within 
the envelope, the void and the surrounding
shell are expected to appear as a region lacking neutral material 
encircled by regions of enhanced \hi\ emission.  
 The analysis of the neutral gas emission distribution in the environs 
of these stars allows identification of such cavities and surrounding shells
associated with the wind sources. 

The criteria adopted to relate an \hi\ cavity and shell to a certain star
are: (i) the star should be located close to the centre of the void or 
within the inner border of the \hi\ surrounding shell; (ii) the ionized 
ring nebula, if present, should appear projected within 
the cavity or close to the inner border of the neutral 
shell; and (iii) the 
kinematical distance to the \hi\ structure should be compatible, within
errors, with the stellar distance. 

To facilitate the visualization of the general characteristics of the \hi\
emission in the line of sight to this region of the Galaxy, 
we show the average \hi\ profile corresponding to an area of 1$^{\circ}$\por 1$^{\circ}$
centered at the position of WR\,85 in the top panel of Figure 2.
The bottom panel shows a plot of the kinematical distance d$_k$ {\it versus} 
the LSR radial velocity for the galactic longitude {\it l} = 347$^{\circ}$, 
as obtained from the circular galactic rotation model by \citet{bb}

Brightness temperatures higher than a few K are observed for velocities 
spanning the range --130 to +50 \kms. According to the quoted circular 
galactic rotation model, gas at negative velocities is placed at kinematical
distances $d_{k}\approx 0-7$ kpc or d$_k \gid 9$ kpc, while positive
velocities are forbidden for distances closer than 17 kpc. Gas within the near 
distance range is most probably related to the Local and the 
Carina-Sagittarius arms \citep{gg}.

To analyze in some detail the neutral atomic gas distribution in the
environs of the WR star, we obtained a series of  \hi\ line images at 
constant velocities, paying particular attention to the gas distribution 
at negative velocities. The analysis of the neutral atomic gas
distribution within the velocity range --150 to 0 \kms\ shows a clear 
void with the star projected close to its centre at velocities of 
about --21 \kms.

\begin{figure}
\includegraphics[angle=0,width=84mm]{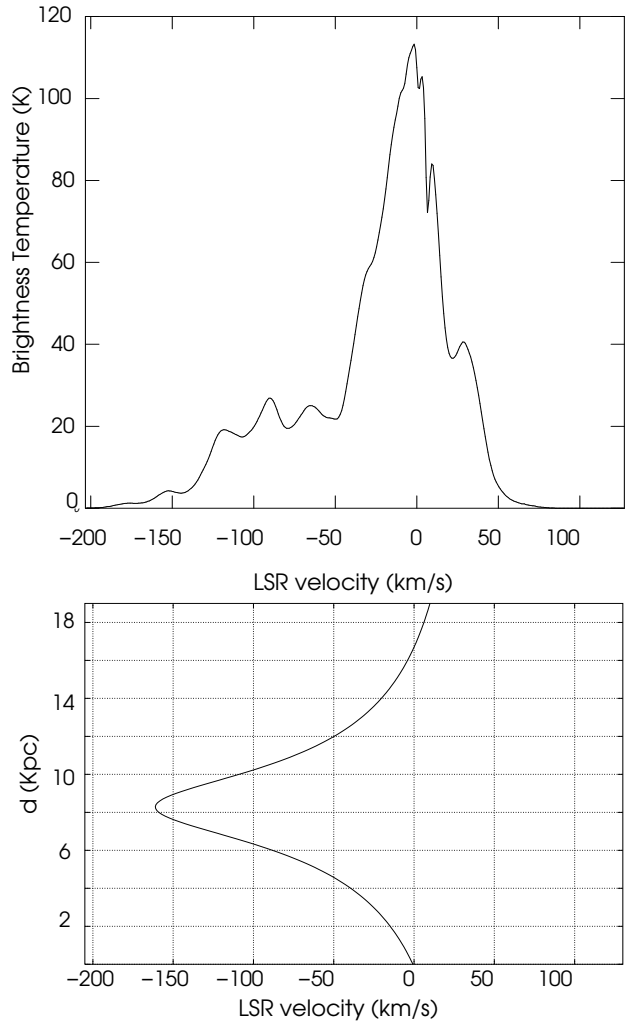}
    \caption{{\it Top panel:} Average \hi\ brightness temperature spectrum 
{\it versus} LSR velocity, corresponding to an area of 1$^{\circ}$\por 1$^{\circ}$ centered at the position of WR\,85.
 {\it Bottom panel:} Kinematical
distance {\it versus} LSR velocity plot obtained from the analytical fit 
to the circular galactic rotation model by \citet{bb} for {\it l} = 
347$^{\circ}$.}
  \label{kdvel}
\end{figure}

\begin{figure}
  \includegraphics[angle=0,width=84mm]{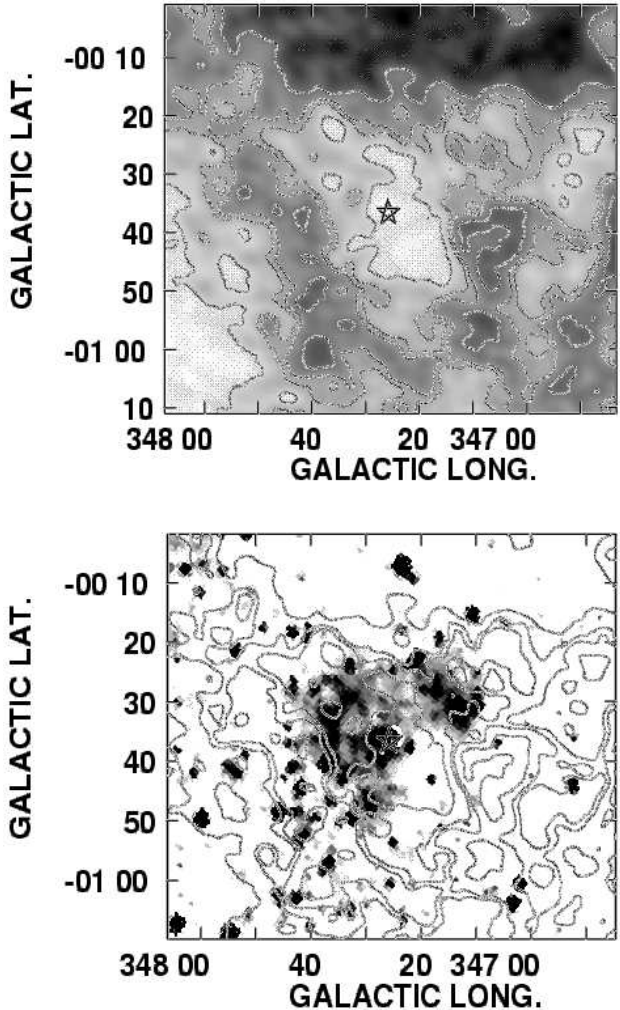}
  \caption{{\it Top panel:} \hi\ column density distribution surrounding 
WR\,85 within the velocity interval --28.0 to --16.5 \kms. The grayscale 
corresponds to (1.2-1.7)$\times$10$^{21}$ cm$^{-2}$ and the contour lines 
are 1.2, 1.4, 1.5 and 1.6$\times$10$^{21}$ cm$^{-2}$. The star symbol 
indicates the position of the WR star. {\it Bottom panel:} Overlay 
of the {\sc shassa} H$\alpha$ image of the nebula and the same \hi\ contours
of the top panel. H$\alpha$ units are arbitrary.}
 \label{NHIwr85}
\end{figure}

\begin{figure}
   \includegraphics[width=84mm]{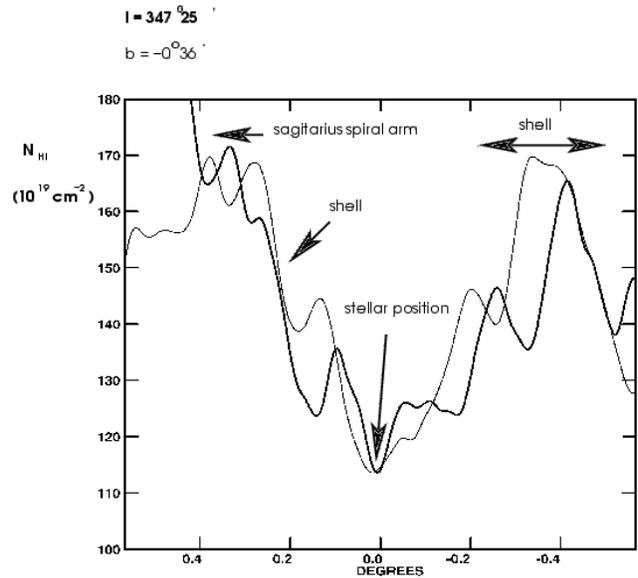}
  \caption{Profiles showing the \hi\ column density  
within the velocity interval --28.0 to --16.5 \kms\ {\it versus}
galactic longitude obtained at {\it b} = --0$^{\circ}$36$^{'}$ 
(thin line) and {\it versus} galactic latitude at {\it l}
 = 347$^{\circ}$25\arcmin\ (thick line). The x-axis is referred to the
stellar position.}

 \label{cortes}
\end{figure}

The top panel of Figure~\ref{NHIwr85} displays the \hi\ column density 
distribution within the velociy interval --28.0 to --16.5 \kms, where the 
void clearly detected.  The void appears surrounded by an almost
circular shell. The centroid of the structure, defined 
following the maxima in the shell, is placed at 
($l,b$) = (347$^{\circ}$26$^{'}$,--0$^{\circ}$37$^{'}$),
close to the position of WR\,85 (indicated by the star symbol).  
The brightness temperature gradient of the structure is slightly steeper 
towards the galactic plane than towards the other sections of the shell,
indicating the presence of higher density regions close to $b$ = 0$^{\circ}$. 

The bottom panel of Fig. 3 displays a superposition of the H$\alpha$ image 
of the region obtained from the  Southern H-alpha Sky Survey Atlas (SHASSA) 
\citep{ga} ({\it greyscale}) and the same \hi\ contours of 
the top panel. 
The 25$^{'}$-diameter outer nebula is clearly detected in the optical 
image (see Fig. 1 for a comparison), while the 9\farcm 3 -diameter inner
 nebula is barely identified.
The bottom panel of Fig. 3 shows that the  25\arcmin -diameter 
optical nebula is projected onto the \hi\
cavity and close to the inner border of the \hi\ shell.  

The systemic velocity of the structure, defined as the velocity 
at which the \hi\ cavity presents its largest dimensions and deepest 
temperature gradient, is v$_{sys} \approx -21$ \kms. This value is compatible, 
within errors, with the velocity of the ionized gas found by \citet{ct}
(--15 \kms).

\begin{figure}
   \includegraphics[width=84mm]{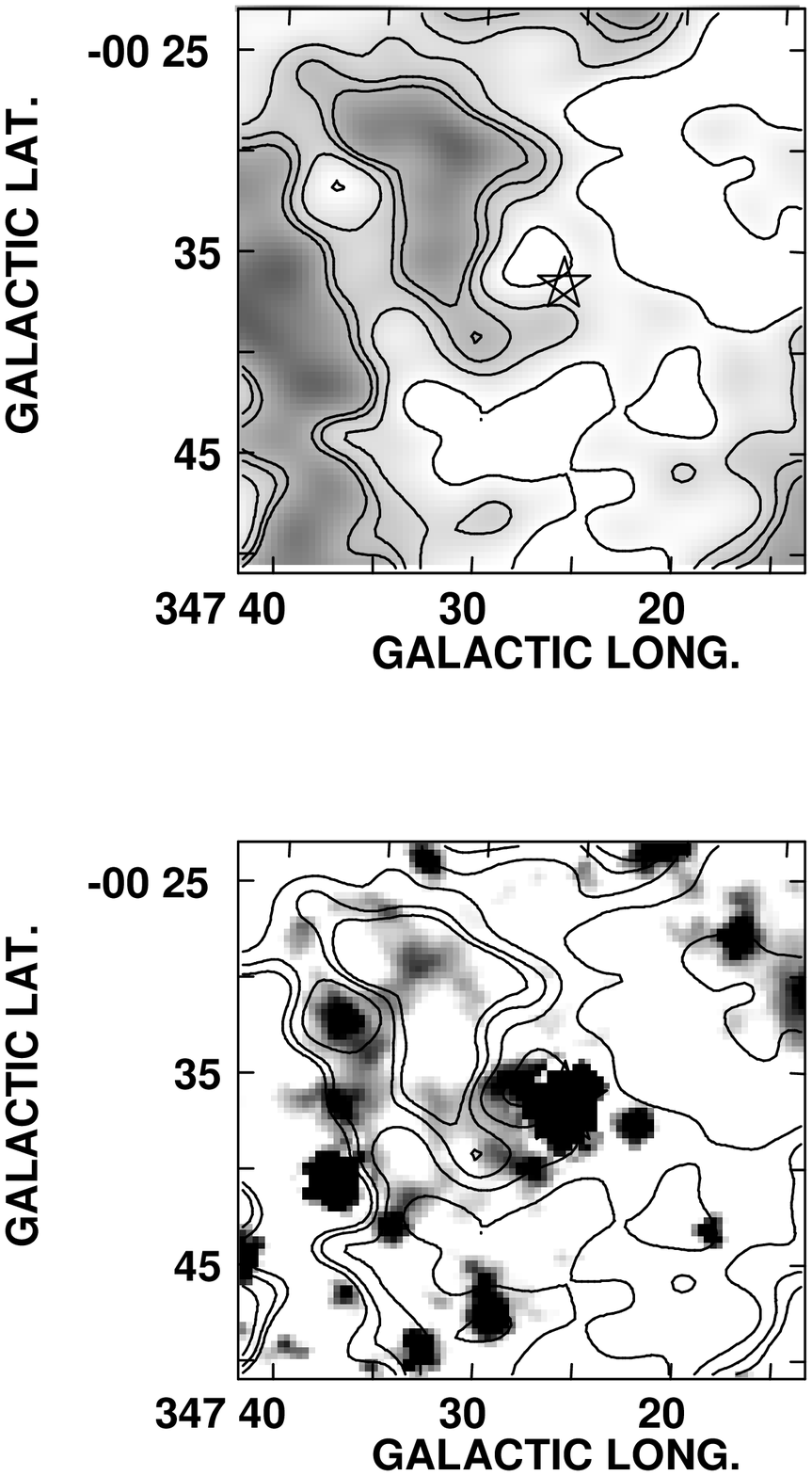}
  \caption{{\it Top panel:} \hi\ emission distribution within  the 
velocity interval --29.3 to --27.6 \kms\ in grey scale and contour lines. 
The grey scale corresponds to 45 to 90 K. The contour lines are
45, 50, 53, 55 K. The cross marks the position of the WR star.
{\it Bottom panel:} Overlay of the {\sc shassa} H$\alpha$ image and the 
same \hi\ contours of the top panel.}

 \label{nueva}
\end{figure}

The presence of the outer optical nebula close to the inner border of 
the \hi\ shell
and the morphological agreement between the RCW\,118 and the \hi\ emission,
along with the agreement between the systemic velocity of the \hi\ structure
and the velocity of the ionized gas indicate that 
the \hi\ feature is related to RCW\,118.    

Figure~\ref{cortes} displays two profiles showing the \hi\ column density
{\it versus} the galactic longitude for $b$ = --0$^{\circ}$36$^{'}$ ({\it 
thin line }) and the galactic latitude for $l$ = 347$^{\circ}$25$^{'}$ 
({\it thick line}). 
This plot clearly shows that the WR star is projected onto a minimun in 
the \hi\ emission distribution. The neutral shell 
is identified with arrows. The wide line shows that 
although most of the neutral gas near $b$ 
= 0$^{\circ}$ is linked to the Sagittarius arm, it is unconnected to the 
\hi\ feature shown in Fig~\ref{NHIwr85}.

 We analyzed is some detail the  \hi\ gas distribution in the close 
environs of the 9\farcm 3 optical ring. The top panel of Fig. 5 
displays the neutral gas emission distribution for the velocity interval 
--29.3 to --27.6 \kms\ (in grey scale and coutour lines), while the 
bottom panel shows an overlay of the optical image (grey scale) and
the \hi\ emission distribution (contour lines). The figure reveals the
presence of neutral gas emission closely bordering the inner semi-circular 
optical ring. The morphological correlation between the \hi\ emission and 
the border of the inner nebula suggests that the ionized and the neutral 
material are related. Some correlation between the inner optical ring 
and neutral hydrogen is also detected at velocities spanning
the range --27.6 to --25.2 \kms. This \hi\ gas is also shown as the
\hi\ peaks interior to the \hi\ shell detected about 10\arcmin\ far from
the star in the profiles in Fig.~\ref{cortes}.  

 The velocity interval at which the correlation between the inner
ring and the \hi\ cloud is better detected suggests 
that the inner optical nebula is located on the approaching section of the 
 expanding shell associated with RCW\,118.

\begin{figure}
   \includegraphics[width=84mm]{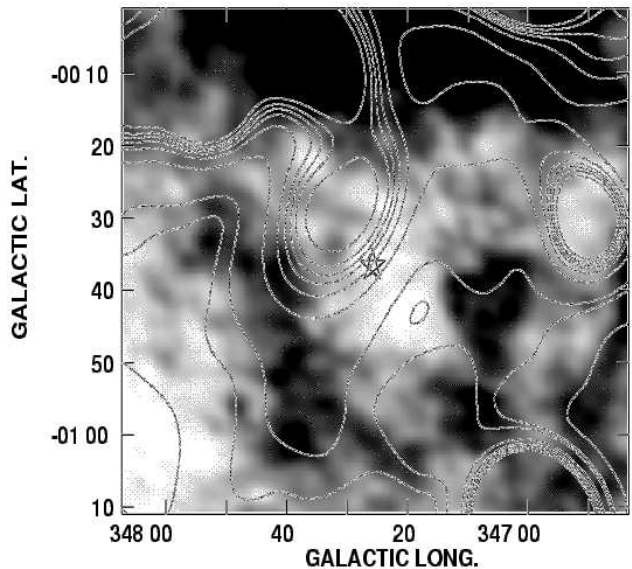}
  \caption{ Overlay of the radio continuum image at 2.4 GHz
 (contour lines) and the \hi\ column density distribution (grayscale).
The star marks the position of WR\,85. Contour lines are 0.04, 0.09,
 0.1, 0.27, 0.29 and 0.32 Jy.}
 \label{2400}
\end{figure}

\subsection{Radio continuum emission}

  Figure \ref{2400} displays an overlay between the radio continuum 
emission at 2.4 GHz (contour lines) and the \hi\ column density distribution
 (greyscale). Within the region of interest, the figure shows a radio 
source centered at ($l,b$) = (347$^{\circ}$32$^{'}$, --0$^{\circ}$29$^{'}$),
coincident with the 
brightest section of RCW\,118 and with the inner optical ring located
close to the star. Weak radio emission can also be detected 
toward higher negative galactic latitudes and lower galactic longitudes, 
coincident with fainter regions of 
RCW\,118. The positional coincidence between both the  radio source 
and the weak radio emission, and RCW\,118 suggests that the emission at 2.4 GHz
originates in the ionized nebular gas.

An image at 4.85 GHz can be obtained from
the PMN Survey (Condon, Broderick \& Seielstadet 1995). However, the presence 
of extended areas lacking radio data within the region of 
interest does not make this image useful. The presence of weak
radio emission in this region is also evident at 4.85 GHz in the survey 
by Haynes, Caswell \& Simonset (1978).

\subsection{Molecular and IR emission} 

The left panel of Figure~\ref{CO} shows the CO(J=1$\rightarrow$0) 
line emission distribution within the velocity range --32.5 to --24.7 
\kms\ taken from the \citet{da} survey. The stellar position is 
indicated by the star. A region of low molecular emission, 
$\sim$ 1$^{\circ}$30\arcmin \por\ 0$^{\circ}$45\arcmin in size,  centered
near (347$^{\circ}$25\arcmin, --0$^{\circ}$ 36\arcmin)  
is evident in  the image.
The right panel of the figure displays an enlargement of the 
molecular void
showing an overlay of the CO emission distribution (thick contour
lines) and the \hi\ column density image of Fig. \ref{NHIwr85} (grayscale 
and thin contour lines). 

 The image on the right shows the clear correspondence between the 
\hi\ structure and the higher galactic longitude section of the CO cavity 
({\it l} $>$ 347$^{\circ}$00$^{'}$). Note that the CO maxima are external 
to the \hi\ shell, suggesting that some stratification in the gas density is 
present. The section of the molecular cavity  at {\it l} 
$<$ 347$^{\circ}$00$^{'}$, which is wider in galactic
latitude, is not linked to the interstellar bubble.

The {\sc hires iras} images at 12, 25, 60 and 100 $\mu$m do not show
any structure connected either to the optical or to the \hi\ and CO 
shells. Only an IR emission gradient probably linked to the galactic plane 
is present in the images at 60 and 100 $\mu$m.

\section{The \hi\ bubble related to WR\,85 and RCW\,118}

\subsection{The distance}

The kinematical distance d$_k$ to the \hi\ structure was estimated from 
the analytical fit to the circular galactic rotation model \citet{bb}. 

This model predicts that gas at v$_{sys}$ = --21 \kms\  should be located
at 2.7$\pm$0.7 kpc or 14$\pm$1 kpc. 
The uncertainties in the quoted values were estimated by assuming the 
presence of non-circular motions of $\approx$ 6 \kms.  The near 
kinematical distance estimate is compatible with the kinematical distance 
of the ionized gas (d$_k$ = 2.0 kpc) derived using the same model.

Bearing in mind the available photometric data for WR\,85
 and the intrinsic color and absolute magnitude corresponding 
to a  WN\,6 star from \citet{vh}, the spectrophotometric distance 
estimate is 
d$_{\*} \approx$ 2.2$\pm$0.9 kpc, where the uncertainty
 in the stellar distance corresponds to the error in the absolute
magnitude ($\pm$0.9 mag). This distance estimate is consistent with
 the near kinematical distance derived from \hi\ data.

\citet{vh} suggests the existence of an OB companion to WR\,85. 
However, the absence of OB absorption lines in the spectrum of WR\,85 
(Gamen, private communication) suggests that the OB companion would be at 
least 2.5 mag weaker than the WR star. Consequently, a correction factor 
to the apparent magnitude of WR\,85 has not been taken into account.

Bearing in mind these values and the distances derived by different  
authors (see $\S1$), we adopted d$_{k}$ = 2.8$\pm$1.1 kpc for 
both the optical and the \hi\ structures. The uncertainty in the 
adopted distance is 40$\%$.

The agreement in position, velocity and distance among the WR star, 
RCW\,118 and the \hi\ feature  suggests that the \hi\ structure is 
associated with the outer optical ring nebula.
The presence of a central star characterized by a strong mass flow 
inside the \hi\ structure and the optical nebula suggests that the 
stellar wind of the WR star and its massive progenitor may have had an 
important role in shaping the nebula.  We therefore interpret the 
\hi\ structure as the neutral gas counterpart of the optical interstellar 
bubble.

\subsection{Main parameters of the interstellar bubble around WR\,85}

\begin{table}
\begin{minipage}{240mm}
\caption{Physical parameters of the HI bubble}
 \begin{tabular}{lc}
\hline
$(l,b)$ centroid     &  347$^{\circ}$26\arcmin,--0$^{\circ}$37\arcmin \\  
Velocity range $v_2,v_1$(\kms)                &   --28.0,--16.5 \\ 
Systemic velocity (\kms)                      &   --21$\pm$5  \\
Expansion velocity (\kms)                     &   9$\pm$2   \\
Angular radius of the \hi\ cavity             &  20\arcmin \\
Angular radius of the \hi\ shell              &  30\arcmin \\  
Angular radius of the \hi\ bubble             &  25\arcmin  \\
Adopted distance (kpc)                        &  2.8$\pm$1.1 \\
Linear radius of the \hi\ bubble (pc)         &  21$\pm$8    \\
\hi\ mass deficiency of the cavity (\msun)    &  220$\pm$170  \\         
\hi\ mass in the shell ($M_{\odot}$)          &  830$\pm$780    \\
\hi\ mass in the \hi\ bubble (\msun)          &  530$\pm$400     \\   
Swept-up neutral mass (\msun)                 &  720$\pm$570      \\
Dynamical age (yr)                            & ($\sim1.1\pm0.5$) \por $10^6$ \\
\hline
\end{tabular}
\end{minipage}
\end{table}

The physical parameters of the \hi\ bubble are summarized  in Table 2. 
The centroid of the structure was defined taking into account the position
of the maxima in the shell. 

The velocity range corresponds to the velocity interval where the structure 
is detected. The systemic velocity was defined in \S 3.1 The expansion 
velocity was estimated as v$_{exp} = (v_2 - v_1$)/2 + 1.6 \kms\ and 
represents a lower limit to the true expansion velocity. The extra 1.6 \kms\ 
allows for the presence of \hi\ in the undetected caps of the expanding shell.
Because of its small column density, these caps are difficult to identify in 
the fore- and background emission.

The cavity was defined following the contour line corresponding to 
1.4\por 10$^{21}$ \cmdos (see Fig.~\ref{NHIwr85}). The angular 
radius of the shell corresponds approximately to the outer border of the 
envelope. It can be
clearly established for {\it b} $<$ --0$^{\circ}$25\arcmin, while confussion
with fore- and background gas precludes the identification of the shell
near {\it b} $<$ --0$^{\circ}$10\arcmin. The size of the  bubble was estimated 
through the maxima in the shell.  

The \hi\ mass deficiency in the cavity and the \hi\ mass in
the shell were obtained from the column density image shown in Fig. 3.
The swept-up neutral mass associated with the \hi\ bubble was derived as a mean
value between the mass deficiency in the cavity and the mass in the
shell assuming a 10\% He abundance. 

Bearing in mind an error of 40\% in the adopted distance, uncertainties 
in radii and masses are about 40\% and 80\%, respectively.     

Evolutionary models of interstellar bubbles allow an estimate of the 
dynamical age as t$_\mathrm{d}$ = 0.50$\times$10$^6 R/v_\mathrm{exp}$ 
yr \citep{mc}, corresponding to the momentum conserving stage of 
an interestellar bubble, where $R$ is the 
radius of the bubble (pc), v$_\mathrm{exp}$ is the expansion velocity (\kms) 
and the constant is the deceleration parameter, which corresponds to a
mean value between the energy and the momentum conserving cases. 
The derived dynamical age {\bf t$_\mathrm{d}$ = 1.1$\times$10$^6$ yr}, is 
larger than the duration of the WN phase of a massive star {(t$_{WN}$ 
= 0.3\por 10$^6$ yr for a rotating star of 40 \msun, \citealt{mm})} and 
suggests that the O-type star 
progenitor of the present WR star has contributed in the formation of the 
bubble.

\begin{figure*}
\includegraphics[angle=270,width=170mm]{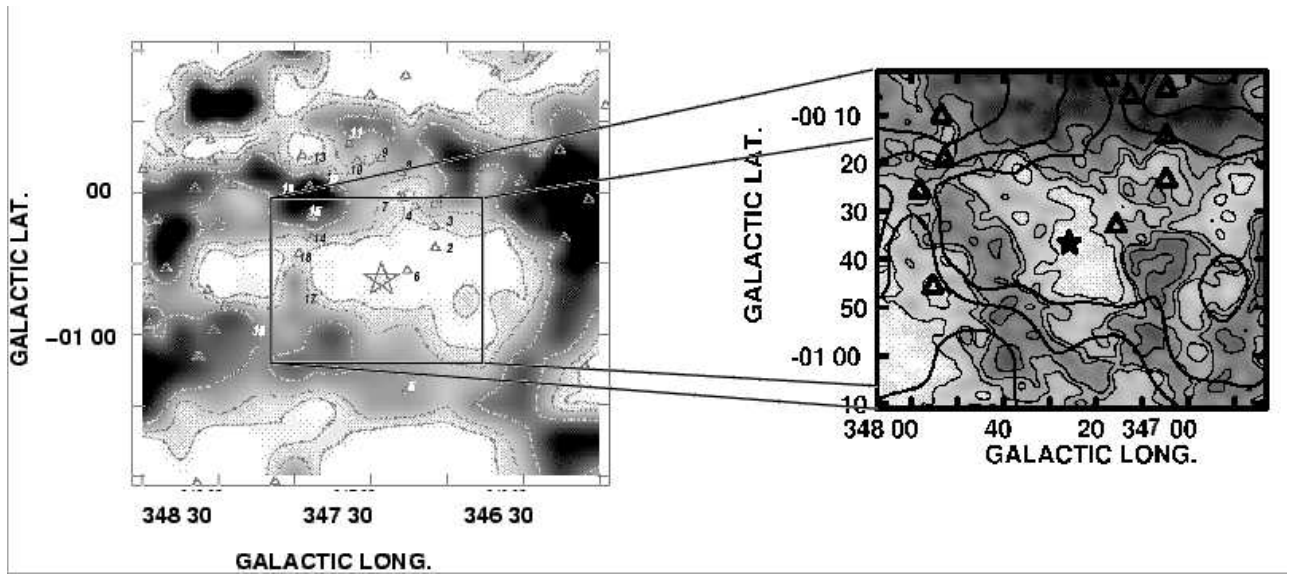}
  \caption{{\it Left panel:}. Mean brightness temperature corresponding 
to the CO emission distribution within the velocity interval --32.5 to --24.7
\kms. The grey scale corresponds to 0.2 to 1.7 K. The contour lines
are 0.3, 0.5 and 0.9 K. The cross marks the position of the WR star. The 
triangles mark the location of {\sc iras} protostellar candidates 
(see text).
{\it Right panel:} Enlargement of the central region 
of the image on the left showing an overlay of the \hi\ ({\it grayscale and 
thin contour lines)} 
and the CO ({\it thick contour lines}) emissions. The star symbol indicates
the stellar position. The triangles mark the location of {\sc iras}
protostellar candidates. }

 \label{CO}
\end{figure*}

A rough estimate of the ionized mass $M_i$ and the electron density $n_e$ 
related to RCW\,118 were obtained from 
the radio continuum image at 2.4 GHz (Fig. 6) using the classical 
expressions by \citet{mh} for the case of a 
spherical \hii\ region.  We derived the flux density by assuming that the 
strong radio source at 347\deg 32\arcmin, --0\deg 29\arcmin\ and part of
the weak radio emission towards higher negative galactic latitudes and
lower galactic longitudes are related to RCW\,118.
For a flux density $S_{2.4}$ = 12 Jy, and 
assuming an electron temperature of 10$^4$ K, and a filling factor $f$ = 
0.35-0.45, we derived $M_i$ = 2100 -- 2400 \msun\ and $n_e$ = 13 -- 15 
cm$^{-3}$.
 The filling factor was estimated from the optical image and 
corresponds to an ionized shell of about 13\arcmin\ in radius 
and 4\arcmin\ in thickness. We assume that about 50-70 $\%$ of the 
shell surface is covered by gas. Uncertainties in the ionized masses and 
electron densities are
about 80$\%$. We have also assumed that He is singly ionized. 
 
Taking into account the neutral atomic and the ionized masses, the total 
mass in the interstellar bubble is $M_s~ \sim$~ 3000 \msun. The average 
ambient density estimated by distributing the total mass within a volume 
of 21 pc in radius is $\sim$3 cm$^{-3}$. 

It is not clear whether part of the molecular material that 
encircles both the \hi\ and \hii\ shells participates in the expansion.
An estimate of the amount of molecular gas can be obtained from
Fig.~\ref{CO} by applying the empirical relation between
W$_{CO}$(=$\int T_{mb}dv$) and the H$_2$ column density N$_{H_2}$ = (1.1$\pm$0.2)\por10$^{20}$
\por W$_{CO}$ \cmdos (K \kms)$^{-1}$, obtained from $\gamma$-ray
studies of molecular clouds in the IV galactic quadrant \citep{sl}.
The total H$_2$ mass is estimated to be 6800 \msun.

\subsection{The energetics}

An estimate of the mechanical luminosity $L_w$ (= \.M V$_w^2$/2) of 
the stellar wind of WR\,85 can be obtained by assuming a typical mass loss 
rate \.M = 2\por 10$^{-5}$ \msunyr\ for the WN phase \citep{ca3}
and a terminal velocity $V_w$ = 1430 \kms\ \citep{rn}. For the previous 
main sequence O-type star phase, we adopted \.M = 2\por 10$^{-6}$ \msunyr\ 
and a terminal velocity $V_w$ = 1000 \kms\ (Prinja, Fullerton \& Crowther 
1996). Mechanical luminosities corresponding to the WR and the O-type star 
phases turn out to be 
$L_{WR}$ = 1.3$\times$10$^{37}$ erg s$^{-1}$ and $L_{O}$ = 
6.3$\times$10$^{35}$ erg s$^{-1}$. Assuming a typical lifetime
 for the WN phase $t_{WN}$ = 0.3\por 10$^6$ yr \citep{mm} and for the 
O-type phase t$_{O}$ = 3\por 10$^6$ yr \citep{cv}, the total stellar wind 
mechanical energy transferred to the ISM  is 
$E_w$= 1.8$\times$10$^{50}$ erg. A similar
value  (1.2\por 10$^{50}$ erg) is derived asuming that the stellar wind of 
the WR star and the previous O-type star phase have blown the gas bubble 
during 0.3\por 10$^6$ and 0.8\por 10$^6$ yr, respectively. These lifetimes 
are compatible with the derived dynamical age. Taking into account the 
large uncertainty in the dynamical age, we believe that the derived 
$E_w$-values can be considered as the lower and upper limits to the true 
stellar wind energy. 
 
The kinetic energy $E_k$ (= M$_s$ V$_{exp}^2$/2) in the interstellar
bubble derived taking into account the expansion velocity
from Table 2 and the swept-up atomic neutral and ionized masses, is 
$E_k$= 8.3$\times$10$^{48}$ erg. If the molecular material also
participates in the expansion, $E_k$= 8.8$\times$10$^{48}$ erg.

The ratio $\epsilon$ (= $E_k$/$E_W$) is in the range 0.005 - 0.03. These
values indicate that the central star is capable 
of blowing the interstellar bubble.
The figure derived for WR\,85 is similar to the ones obtained for most of
the \hi\ interstellar bubbles found around  O- and WR-stars.
These values support the interpretation that bearing in mind the standard
energy conserving model by \citet{we}, the observed stellar wind 
energy appears to be too high for the observed bubble dynamics, as
pointed out by \citet{co}, and suggest that the bubbles are most
probably in the momentum conserving stage. 
However, the drain of energy from the bubble through the patchy envelope
can not be ruled out.

The ambient density obtained by distributing the total mass within the 
volume of the bubble (see \S 4.2) is compatible with the value derived 
for the momentum conserving case ($\sim$ 3 cm$^{-3}$).
 
The SNR G\,347.3-0.5 is seen projected onto the same area. The derived
 distance to the SNR, $d$ = 6 kpc, based on molecular line information 
\citep{sl}, precludes any relation between the SNR
and RCW\,118 and the \hi\ shell.

\section{Conditions for star formation}

The high velocity mass flow produces a drastic change in the physical
conditions of the surrounding ISM. Shock fronts linked to stellar winds
from massive stars may induce star formation in the high density regions
of the neutral shells, where material has accumulated and conditions for
star formation may have been favoured. Then, it is important to analyze the 
presence of star formation indicators in the neutral shells. 

 Since protostellar candidates can be identified as infrared sources 
in the {\sc iras} Point Source Catalogue, we performed a search for 
{\sc iras} point sources projected onto a region of 3\deg\ in size 
centered at the stellar position whose 
energy distributions are compatible with protostellar objects according to 
the criteria listed by Junkes, F\"urst \& Reich (1992).  The IR sources
found in our search are indicated by triangles in Fig.~\ref{CO} (left panel). 
Only the 
ones within an area of 1\deg\ in radius centered at the WR position 
(triangles with numbers) are listed in Table 3, which shows the source number,
 its {\sc iras} 
number identification, and the fluxes in the four {\sc iras} bands.

 It is clear from the left panel of the figure that most of the 
sources are projected onto regions of strong CO emission or close to the 
galactic plane. The triangles in the right panel of Fig.~\ref{CO} mark the 
position of {\sc iras} point sources.  This panel shows the close
correspondence between the IR sources 2, 3, 6, 14, 15, 17 and 18, and the 
molecular and \hi\ shells linked to RCW\,118, suggesting that sequential 
star formation may be occuring in some sections of the neutral envelope, 
where conditions for stellar formation might have been developed. 
 We note that the distances to these sources are unknown.

\begin{table*}
\begin{minipage}{84mm}
\caption{Protostellar candidates within a region of 1$^\circ$ in radius 
centered on WR\,85}
\begin{tabular}{lcrccccc}
\hline\hline
\#  &  $l$   &  $b$  &  {\sc iras} & F(12$\mu$m) & F(25$\mu$m) & F(60$\mu$m) & F(100$\mu$m) \\
    &        &       & designation       \\
    &  ($^{\circ}$ $^{'}$) &  ($^{\circ}$ $^{'}$ )    &  (Jy)   &   (Jy)      &  (Jy)       & (Jy)        &  (Jy) \\
\hline
1  &  347 4 & -0  4 & 17076-3940 &  3.75   & 3.82        &   131.      &  410.  \\ 
2  &  347 4 & -0  22& 17089-3951 &  4.38   & 13.         &   98.5      &  234. \\ 
3  &  347 4 & -0  13& 17083-3946 &  7.94   & 4.42        &     39.70     &  182. \\ 
4  &  347  12& -0  6 & 17081-3935 & 2.75   & 4.13        &     94.5      &  435. \\ 
5  &  347  15&-1   22& 17137-4018 &  18.89  & 7.55        &          63.5       & 188. \\ 
6  &  347  15&-0   33& 17101-3948 & 3.25  &  4.57       &       18.70      & 64.59 \\ 
7  &  347  17&-0   1 & 17081-3928 & 11.    & 74.30       &         221.       & 739. \\ 
8  &  347  17& 0   8 & 17074-3922 &  3.78   & 7.38        &       256.       & 1120. \\ 
9  &  347  25& 0   13& 17074-3911 &  4.69   & 2.42        &         150.       & 397. \\ 
10 &  347  34& 0   12& 17079-3905 &   57.20  & 396.        &     4030.      & 8220. \\ 
11 &  347  38& 0   20& 17076-3858 &   2.67   & 4.03        &     80.80      & 280. \\ 
12 &  347  43& 0   1 & 17091-3905 &   62.20  & 24.89       &        37.5       & 211. \\ 
13 &  347  45& 0   9 & 17087-3859 & 4.00.  & 25.10       &      245.       & 628. \\ 
14 &  347  52&-0   18& 17110-3910 &     3.43   & 18.29       &    221.       & 941. \\ 
15 &  347  52&-0   10& 17105-3904 &    1.60   & 9.10        &      91.69      & 226. \\ 
16 &  347  53& 0   2 & 17096-3856 &     24.00  & 164.00      &  1050.      & 1720. \\ 
17 &  347  55&-0   45& 17130-3923 &  3.49     & 21.90      &         145.       & 148. \\ 
18 &  347  58&-0   25& 17118-3909 &    9.72   & 80.30       &   1120.      & 3000. \\ 
19 &  348  13&-0   58& 17149-3916 &    209.   &  988.       & 6760.      & 9160. \\ 
\hline
\end{tabular}
\end{minipage}
\end{table*}

\section{Summary}

We have analyzed 21cm line data belonging to the Southern Galactic Plane
Survey searching for an \hi\ bubble related to WR\,85 and its ring nebula
RCW\,118. The main results of our search can be summarized as follows:
\begin{enumerate}
\item[(i)] The \hi\ data allowed the identification of a cavity in the neutral
gas distribution surrounded by a slowly expanding shell. The structure
was detected at velocities spanning the interval 
--28.0 to --16.5 \kms, and has a systemic velocity of --21$\pm$5 \kms. The cavity
can also be identified in the CO emission distribution at similar
velocities.\
\item[(ii)] The ring nebula RCW\,118 appears projected onto the inner border of the
\hi\ shell. \
\item[(iii)] The systemic velocity of the \hi\ structure is similar to the
velocity of the ionized gas in RCW\,118.\
\item[(iv)] WR\,85 is seen in projection onto the central part of the \hi\ cavity 
and close to its geometrical center.\
\item[(v)] The coincidence, within errors, of the kinematical distance to the 
\hi\ structure and the stellar distance strongly suggests that the WR star
is placed within the cavity.
\item[(vi)] The morphological coincidence between the inner optical 
nebula and the \hi\ emission at about --28 \kms\ suggests that the 
9\farcm 3 nebula is located on the approaching part of the expanding shell.
The WR star, which appears closely related to the inner nebula, is also 
probably located near the approaching part of the shell.
\item[(vii)] The CO emission distribution reveals the presence of a molecular
 shell almost encircling the HI envelope.
\item[(viii)] The position of the {\sc iras} protostellar candidates 
 projected onto the molecular shell, suggests that star formation 
might be occuring in the
region. However, additional studies should be performed to investigate
this point.
\item[(ix)] The stellar wind of WR\,85 is capable of blowing the observed ionized 
and neutral structure.
\end{enumerate}

The bulk of evidence (i) to (ix) strongly indicates that the \hi\ 
structure is the neutral gas counterpart of the optical bubble and
shows the action of the stellar winds on the surrounding material.

Adopting a distance of 2.8$\pm$1.1 kpc, the neutral interstelar bubble 
has a linear radius of $\sim$21 pc. Taking into account an expansion velocity 
of $9\pm$2 \kms, its dynamical age is $1.1\times 10^6$ yr, suggesting that 
both the present WR star and its massive stellar progenitor have contributed
in shaping the bubble.  The associated neutral and ionized masses are 
3000 M$_{\odot}$, which indicates that the bubble evolved in a medium with 
an average ambient density of 3 cm$^{-3}$.

\section*{Acknowledgments}

 We acknowedge the referee Dr A. P. Marston for many suggestions that 
improved the final presentation of this paper.
We thank Dr. T. Dame for making available to us the CO data, and Dr. R. 
Gamen for allowing us to use information on WR\,85 in advance of publication.
This project was partially financed by the Consejo Nacional de 
Investigaciones Cient\'{\i}ficas y T\'ecnicas (CONICET) of Argentina under 
project PIP 607/98 and FCAG, UNLP under projects 
11/G049 and 11/G072, and Agencia Nacional de Promoci\'on Cienf\'{\i}fica y Tecnol\'ogica (ANPCYT) under project PICT 14018/03.
The Digitized Sky Survey (DSS) was produced at the Space Telescope Science
Institute under US Government grant NAGW-2166.

\end{document}